\documentclass[a4paper,11pt]{article}
\usepackage{pos}

\title{Negative Coupling $\phi^4$ on the Lattice}

\author*{Paul Romatschke}
\affiliation{Department of Physics, University of Colorado, Boulder, Colorado 80309, USA}
\affiliation{Center for Theory of Quantum Matter, University of Colorado, Boulder, Colorado 80309, USA}

\emailAdd{paul.romatschke@colorado.edu}

\abstract{Triviality of $\phi^4$ theory in four dimensions can be avoided if the bare coupling constant is negative in the UV. Theories with negative coupling can be put on the lattice if the integration domain for $\phi(x)$ is contour-deformed from the real to the complex domain. In 0+1d (quantum mechanics), one can recover results from $\mathcal{PT}$-symmetric quantum mechanics in this way. In this work, I report on an attempt to put negative coupling $\phi^4$ theory in 4 dimensions on the lattice.}

\FullConference{The 40th International Symposium on Lattice Field Theory (Lattice 2023)\\
July 31st - August 4th, 2023\\
Fermi National Accelerator Laboratory\\}

\begin{document}
\maketitle

\section{Introduction}

It is commonly assumed that scalar field theory with quartic interaction in four dimensions is quantum trivial in the continuum, meaning that all expectation values of the field theory are Gaussian. Many people in the lattice community have worked on this subject \cite{Wilson:1973jj,Frohlich:1982tw,Luscher:1987ay,Heller:1993yv}, essentially always corroborating the notion of quantum triviality in scalar field theory.

In 2019 quantum triviality for one and two component scalar fields has been rigorously proved in Ref.~\cite{Aizenman:2019yuo}. The mathematical proof rests on certain assumptions regarding the defining action of the field theory, in particular that the interaction potential is bounded from below. In different words, the proof assumes that the lattice scalar self-coupling is positive definite $\lambda_0>0$.

However, exact analytic calculations for interacting scalars, which are possible in the O(N) model in the large N limit \cite{Linde:1976qh,Abbott:1975bn,Romatschke:2022jqg,Romatschke:2022llf}, suggest that the running coupling in the continuum approaches zero from below \cite{Romatschke:2023sce}
\begin{equation}
\lim_{a\rightarrow 0}\lambda_0\rightarrow 0^-\,,
\end{equation}
where $a$ is the lattice spacing for the theory.

From the outset, negative coupling lattice field theory faces an important issue: The standard path integral representation for the partition function can not be used to access the negative coupling region, because the potential is unbounded from below. This is in complete analogy to the integral representation of many special functions in mathematics, such as the $\Gamma(x)$ function and the Riemann  $\zeta(x)$ function, which have integral representations only for $x\in \mathbb{R}^+$. However, analytic continuations of these functions to the whole complex plane, and in particular $x\in \mathbb{R}^-$ have been known by mathematicians for centuries.

In  the context of quantum mechanics, the use of analytic continuation to access Hamiltonians with classically unbounded potentials is relatively recent, cf. Ref.~\cite{Bender:1998ke}. On the lattice, scalar field theory with negative (and even complex) coupling on the lattice in 1d and 2d was considered in a pioneering study by Lawrence, Oh and Yamauchi in Ref.~\cite{Lawrence:2022afv}. In this study, the authors used contour deformations of the lattice partition function as a practical method to implement analytic continuation. The 1d case was further studied in comparison to Hamiltonian spectra in Ref.~\cite{Lawrence:2023woz}, where the relation between different analytic continuations and ${\cal PT}$-symmetry was further clarified. 

Studies of scalar field theory with negative coupling in 4d on the lattice are even more recent, in particular for single component scalar fields studied in Ref.~\cite{Romatschke:2023sce} and proposals for lattice actions for multi-component scalar fields given in Ref.~\cite{Weller:2023jhc}. The present write-up outlines the setup for 4d scalar field theory with negative coupling on the lattice.

\section{A toy model in 0d}

\begin{figure}
  \centering
  \includegraphics[width=.5\linewidth]{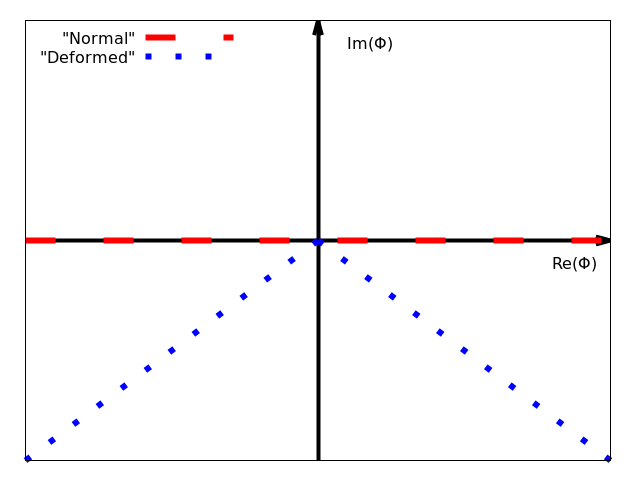}
  \caption{Contour deformation for integrations in the complex domain of $\phi(x)$. ``Normal'' integration for $\phi(x)$ is for along the real line, but the ``Deformed'' integration contour is along two segments at an angle to the real line.  \label{fig1}}
\end{figure}

The simplest toy model for this case is 0d field theory with a partition function given by the integral representation
\begin{equation}
  \label{z1}
Z^0(\lambda)=\int_{-\infty}^\infty d\phi\, e^{-\lambda \phi^4}\,,\quad {\rm Re}(\lambda)>0\,.
\end{equation}
This integral representation can not be used to access values of $Z^0(\lambda)$ for $\lambda \in \mathbb{R}^-$. However, one may define an analytic continuation of $Z^0(\lambda)$ that lets us access $\lambda<0$. Specifically, instead of integrating along $x\in \mathbb{R}$, consider for the negative coupling partition function the ``contour-deformed'' integral
\begin{equation}
  \label{ZC}
Z^0_{\cal C}(g=-\lambda=)=\int_{\cal C} d\phi\, e^{g \phi^4}=\int_0^\infty ds e^{-i\alpha} e^{g s^4 e^{-4 i \alpha}}+\int_{-\infty}^0 ds e^{i\alpha} e^{g s^4 e^{4 i \alpha}}\,,
\end{equation}
if ${\cal C}$ is given by a path in the complex plane sketched in Fig.~\ref{fig1}, or analytically given by
\begin{equation}
  \label{contourdef}
\phi\rightarrow s\left[e^{i \alpha}\theta(-s)+e^{-i \alpha}\theta(s)\right]\,,\quad s,\alpha \in \mathbb{R}\,,
\end{equation}
with $\alpha\sim \frac{\pi}{4}$ the angle with respect to the real axis. The precise value of $\alpha$ is unimportant, as long as ${\rm Re}\left(e^{\pm 4i\alpha}\right)<0$, because then the integrals in (\ref{ZC}) are convergent and can be evaluated to give
\begin{equation}
  \label{ZC2}
Z^0_{\cal C}(g)= \Gamma\left(\frac{5}{4}\right)g^{-\frac{1}{4}} \left(e^{-i\alpha} \left(e^{-4 i \alpha- i \pi}\right)^{-\frac{1}{4}}+ e^{i\alpha} \left(e^{4 i \alpha- i \pi}\right)^{-\frac{1}{4}} \right)=\sqrt{2}\Gamma\left(\frac{5}{4}\right)g^{-\frac{1}{4}}\,.
\end{equation}
For the simple toy model at hand, one can recognize $Z^0_{\cal C}(g)={\rm Re}\,Z^0(\lambda=-g)$ to be related to (\ref{z1}) when \textit{first performing the integral for $\lambda>0$}, then \textit{using the analytic continuation of the root function for $\lambda\rightarrow -g<0$}, and finally taking the real part.

\section{Quantum mechanics: 1d}

A slightly more interesting toy model is provided by considering the partition function for quantum mechanics, given by
\begin{equation}
Z^1(\lambda)=\int {\cal D}\phi e^{-S_E}\,,\quad S_E=\int_0^\beta d\tau\left[\frac{1}{2}\dot\phi^2+\lambda \phi^4\right]\,,
\end{equation}
where $\beta$ is the radius of the Euclidean circle.
As in the 0d toy model, this integral representation exists for $\lambda>0$, and direct evaluation of $Z^1(\lambda)$ for $\lambda=-g<0$ is not possible. However, on the lattice with $N$ sites one can employ the same contour deformation (\ref{contourdef}) at every lattice site in order to define a contour-deformed partition function for the negative coupling theory
\begin{equation}
  \label{Z1C}
Z^1_{\cal C}(g=-\lambda)=\int \prod_{i=1}^N \frac{d\phi_i}{\sqrt{2\pi}} e^{-\sum_i \left(\frac{N}{2}\left(\phi_{i+1}-\phi_i\right)^2+\frac{g \beta^3}{N} s_i^4\right)}\,,\quad \phi_{N+1}=\phi_1\,,
\end{equation}
where for simplicity I've taken the angle $\alpha=\frac{\pi}{4}$ and rescaled $\phi\rightarrow \sqrt{\beta} \phi$. Note that this mixed form with both (complex-valued) $\phi_i$'s and (real-valued) $s_i$'s allows for compact notation of $Z^1_{\cal C}$, but belies the complicated structure of the integral\footnote{Note that for quantum mechanics, a different choice of contour ${\cal C}$ allows one to bring the action into an alternative Hermitian form \cite{Jones:2006qs}, which greatly simplifies the study of negative coupling quantum mechanics.}. In particular, the discretized derivative term $\left(\phi_{i+1}-\phi_i\right)$ mixes the left part of the integration contour (\ref{fig1}) on site $i$ with the right part on site $i+1$, with the appropriate weights $e^{\pm i \alpha}$ in both the exponent and the Jacobian. Clearly, the action is complex, and therefore the theory possesses a sign problem. However, because the highest polynomial term in the action is proportional to $g s_i^4$, $Z^1_{\cal C}(g)$ constitutes a convergent integral representation for $g>0$ on the lattice.

\begin{figure}
  \centering
  \includegraphics[width=.5\linewidth]{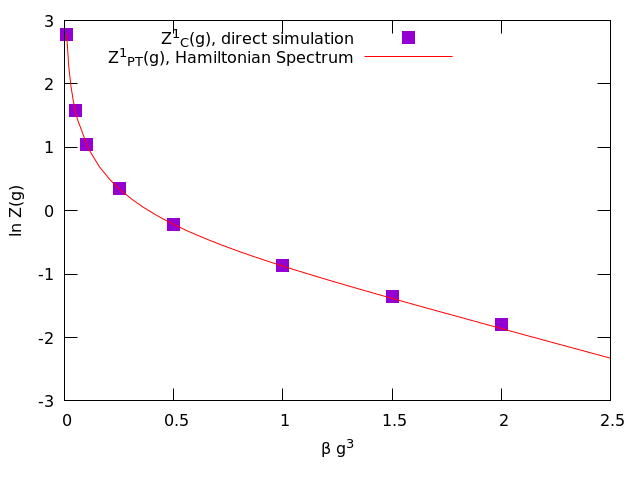}
  \caption{Quantum mechanical partition function for ``upside-down'' quartic potential $V(x)=-g x^4$, once from contour deformation (\ref{Z1C}) for $N=5$ sites and once from the spectrum of the ${\cal PT}$-symmetric Hamiltonian (\ref{Z1pt}). \label{fig2}.}
\end{figure}

Evaluation of $Z^1_{\cal C}(g)$ is possible for a small number of sites N by direct integration \cite{Lawrence:2023woz}, see figure \ref{fig2}. The resulting partition function is real and positive-definite. Moreover, it can be compared to the partition function for the ${\cal PT}$-symmetric theory
\begin{equation}
  \label{Z1pt}
  Z_{\cal PT}^1(g)=\sum_{n=0}^\infty e^{-\beta g^{\frac{1}{3}}E_n}\,,
\end{equation}
with $E_n$ the energy-levels calculated in Ref.~\cite{Bender:1998ke}. In the same units as used in the rest of this work, the first few energy levels are given by $E_0\simeq 0.930546$, $E_1\simeq 3.781896$, $E_2\simeq 7.435067$, $E_3\simeq 11.628327$ and higher-lying eigenvalues by the WKB formula $E_n=\left[\frac{3 \sqrt{2}\left(n+\frac{1}{2}\right)}{\Gamma^2\left(\frac{1}{4}\right)}\right]^{\frac{4}{3}}\pi^2$. A comparison of $Z^1_{\rm cal PT}(g)$ and $Z^1_{\cal C}(g)$ is shown in Figure \ref{fig2}. As can be seen from this figure, the contour-deformed partition function agrees with high numerical accuracy with the partition function obtained from the Hamiltonian spectrum of the ${\cal PT}$-symmetric theory. I take this as further evidence that the contour-deformed lattice theory can be used to access properties of the negative-coupling system.

\section{Negative Coupling $\phi^4$ theory in 4d}

Motivated by the successes in the lower-dimensional toy models, I consider the contour-deformed partition function of scalar $\phi^4$ theory in 4 dimensions:
\begin{equation}
  \label{z4}
Z_{\cal C}^4(g)=\int_{\cal C} \prod_{{\rm sites}\ i} \frac{d\phi_i}{\sqrt{2\pi}} e^{-S_L}\,,\quad S_L=\sum_{{\rm sites}\ i} \left[\sum_{\hat{x}}\frac{\left(\phi_{i+\hat{x}}-\phi_{i}\right)^2}{2}+\frac{m^2 \phi_i^2}{2}-g \phi_{i}^4\right]\,,
\end{equation}
where I consider a lattice with $N_\tau\times N_\sigma^3$ sites and $\sum_{\hat{x}}$ is over all Euclidean directions $\tau,x,y,z$. For each site, I choose the contour (\ref{contourdef}) with $\alpha=\frac{\pi}{4}$ so that in particular
\begin{equation}
-g \phi_i^4\rightarrow +g s_i^4\,,
\end{equation}
is the highest polynomial power in the action $S_L$. Integration over $s_i$ therefore is well-defined at every site, but the action of the theory is neither real-valued nor polynomial in the continuum limit. Yet it constitutes a well-defined lattice theory that is amenable to numerical integration.

The fact that the action is complex prohibits the naive use of Monte Carlo importance sampling techniques to calculate $Z^4_{\cal C}(g)$, because the theory has a sign problem. By contrast, for a small number of sites, $Z^4_{\cal C}(g)$ is amenable to direct numerical integration. In practice, direct numerical integration techniques on modern multi-core computers are limited to oscillatory integrals of up to (roughly) 30 dimensions. Therefore, evaluation of $Z_{\cal C}^4(g)$ via direct numerical integration is limited to tiny lattices $N_\sigma^3\times N_\tau$ of sizes $2^3\times 1, 2^4$ and $3^3\times 1$. Because of this limitation, I report on results for $Z_{\cal C}^4(g)$ for $N_\tau=1$ and $N_\sigma=2,3$ only.

While non-perturbative results for single-component scalar field theory in four dimensions are notoriously hard to obtain, exact results exist for the O(N) model in the large N limit. The results obtained in Refs.~\cite{Romatschke:2022jqg,Romatschke:2022llf,Romatschke:2023sce,Weller:2023jhc} indicate that in the large N limit, the theory is asymptotically free with a second order phase transition at a critical temperature $T_c$. Even though applying the insights gained in the $N\rightarrow \infty$ limit to the case of single component scalars with N=1 may be challenging, one can nevertheless try to use the large N results as a qualitative guidance. In particular, for this work the relation between the bare lattice coupling $g$ and the lattice spacing $a$ is assumed to be given by the analytic O(N$\gg$1) result
\begin{equation}
g=-\frac{2\pi^2}{\ln\left(T_c a\right)}\,,
\end{equation}
with $T_c$ again the critical temperature of the theory. Using the temperature in lattice units
\begin{equation}
  T=\frac{1}{N_\tau a}\,,
\end{equation}
allows to relate the bare coupling constant to the physical temperature in units of $T_c$ as
\begin{equation}
  \label{rung}
g=\frac{2\pi^2}{\ln\left(\frac{N_\tau T}{T_c}\right)}\,.
\end{equation}

In practice, direct evaluation of (\ref{z4}) is performed for the choice
\begin{equation}
m^2=1\,,
\end{equation}
in lattice units and various values of the bare lattice coupling $g$. From (\ref{rung}), one expects $\lim_{T\rightarrow \infty} g\rightarrow 0$ in the high temperature limit. By contrast, large values of $g$ are expected to probe the low-temperature properties of the theory.

\begin{figure}
  \centering
  \includegraphics[width=.5\linewidth]{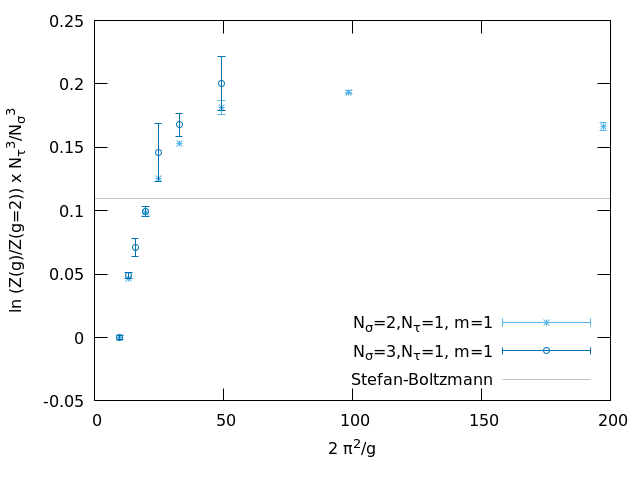}
  \caption{Reduced partition function (\ref{red}) for negative coupling $\phi^4$ theory on lattices $N_\tau=1,N_\sigma=2,3$ via direct numerical integration of (\ref{z4}). For reference, also the Stefan Boltzmann limiting value $\frac{\pi^2}{90}$ is shown. \label{fig:three}}
\end{figure}

In order to compare different lattice simulations, I arbitrarily fix $Z_{\cal C}^4(g=2)$  to correspond to the partition function value at zero temperature. Using this choice to subtract the zero-temperature contribution, the combination
\begin{equation}
  \label{red}
\frac{N_\tau^3}{N_\sigma^3}\ln \frac{Z_{\cal C}^4(g)}{Z_{\cal C}^4(g=2)}\simeq \frac{p(T)-p(T=0)}{T^4}\,,
\end{equation}
becomes a measure of the reduced finite temperature pressure $p(T)$, or equivalently the number of degrees of freedom of the theory. Results for this quantity are shown in Fig.~\ref{fig:three} in comparison to the expected continuum high-temperature Stefan Boltzmann limit. From Fig.~\ref{fig:three}, one can see that there is little difference between the $N_\sigma=2,3$ results. In addition, the obtained shape of the reduced partition function (\ref{red}) matches the expectation for the reduced pressure of the O(N) model studied in Ref.~\cite{Romatschke:2022jqg}. More simulation data, in particular for $N_\tau>1$ would be needed in order to offer more conclusive interpretations.

\section{Conclusions}

In this work, I reported on attempts to study negative coupling $\phi^4$ theory on the lattice using contour deformation techniques. In 4d, the contour-deformed theory leads to a complex but stable lattice action that is amenable to direct numerical integration techniques. I performed such direct numerical integrations on small lattices $N_\tau=1,N_\sigma=2,3$ finding encouraging results.

\acknowledgments{I am thankful to Uli Wolff for comments and to Scott Lawrence for helpful discussions. This work was supported by the Department of Energy, DOE award No DE-SC0017905.}

\bibliographystyle{JHEP}
\bibliography{PT}

\end{document}